\documentclass[12pt,thmsa,a4paper]{article}

\usepackage{amsmath}
\usepackage{amssymb}
\usepackage{epsfig}
\usepackage[latin1]{inputenc}

\newcommand{\p}{\partial}
\newcommand{\n}{\nabla}

\newcommand{\Sss}{\left(1-\frac{2M}{r}\right)}
\newcommand{\Th}{T^{\theta}{}_{\theta}}
\newcommand{\E}[1]{\left< #1 \right>}

\newcommand{\St}[1]{\left| #1 \right>}
\newcommand{\U}[1]{\left<U\left| #1 \right|U\right>}

\newcommand{\vp}{\varphi}
\newcommand{\ve}{\varepsilon}
\newcommand{\1}{1\hspace{-0.243em}\text{l}}

\begin{document}

\begin{titlepage}
\renewcommand{\thefootnote}{\fnsymbol{footnote}}

\hfill TUW--04--20 \\

\begin{center}

{\Large\bf Effective Action and Hawking Flux from Covariant
Perturbation Theory}\\
  \vspace{7ex}
  D.~Hofmann\footnotemark[1],
  W.\ Kummer\footnotemark[2],
  \vspace{7ex}

  {\footnotemark[1]\footnotemark[2]\footnotesize Institut f\"ur
    Theoretische Physik \\ Technische Universit\"at Wien \\ Wiedner
    Hauptstr.  8--10, A-1040 Wien, Austria}
  \vspace{2ex}

   \footnotetext[1]{E-mail: \texttt{hofmann@hep.itp.tuwien.ac.at}}
   \footnotetext[2]{E-mail: \texttt{wkummer@tph.tuwien.ac.at}}
\end{center}
\vspace{7ex}
\begin{abstract}
The computation of the radiation flux related to the Hawking
temperature of a Schwarzschild Black Hole or
another geometric background is still well-known to be fraught with a
number of delicate problems. In spherical reduction, as shown by one
of the present authors (W. K.) with D.V. Vassilevich, the correct
black body radiation
follows when two ``basic components'' (conformal anomaly and a
``dilaton'' anomaly) are used as input in the integrated
energy-momentum conservation equation. The main new element in the present
work is the use of a quite different method, the covariant perturbation theory
of Barvinsky and Vilkovisky, to establish directly the full effective
action which determines these basic components.
In the derivation of W. K. and D.V. Vassilevich the computation of the
dilaton anomaly implied one potentially doubtful intermediate step
which can be avoided here. Moreover, the present approach also is
sensitive to IR (renormalisation) effects. We realize that
the effective action naturally leads to expectation values in the Boulware
vacuum which, making use of the conservation equation, suffice
for the computation of the Hawking flux in other quantum states,
in particular for the relevant Unruh state. Thus, a rather comprehensive
discussion of the effects of (UV \emph{and} IR) renormalisation upon
radiation flux and energy density is possible.
\end{abstract}


\vfill

\end{titlepage}

\section{Introduction}

Almost three decades after the (theoretical) discovery of quantum
radiation from the event horizon of a geometrical background, in particular
from a Black Hole (BH) \cite{haw75,unr76}, somewhat
surprisingly, still the existence of open problems is an acknowledged
phenomenon -- even when only large BH-s are considered which, to a
good approximation, represent a time-independent curved background.

Actually, the computation of the Hawking effect does not require
an analysis of the complete evolution of the (massless) fields between
infinitely early and infinitely late times. It is sufficient to
exploit the Energy-Momentum (EM) tensor near the future horizon only.
The relation between Hawking temperature and the radiation
flux at infinity is still the object of some debate. The activity
in this field has been rekindled by the work of Bousso and Hawking
\cite{boh97} who, on the basis of a computation in 2d dilaton gravity
resulting from spherically
reducing Einstein gravity \cite{tih84}, claimed that an incoming asymptotic
flux and thus {}``anti-evaporation'' occurs. Actually already some
time before ref. \cite{boh97} it had been argued \cite{mwz94}
that the so-called conformal anomaly $\E{T}_2$,
the trace of the EM tensor in two dimensions\footnote{A lower
index $2,4$ attached to an expectation value means computation in
spherically reduced gravity or directly in four dimensions,
respectively.}, which had been the
only input considered\footnote{The correct expression for the
conformal anomaly in the presence of a $2d$ dilaton field for
spherically reduced gravity \cite{mwz94}
and for general $2d$ dilaton theories \cite{klv97b,klv98} taken alone
yields the same unphysical flux. For a comprehensive review of general
$2d$ dilaton models we refer to \cite{gkv02}.} in ref. \cite{boh97},
cannot provide the complete answer.

The problem has been focused in refs. \cite{baf98,baf99,lmr99}
where the relation
between the missing second piece, the 2d ``dilaton anomaly'' $\E{\Th}_2$,
to the pressure component $\E{T^{\theta}{}_{\theta}}_4$ of the EM tensor
in the original 4d theory (in coordinates time, radius
and angles $\theta, \vp$) has been established. In the following
these two essential ingredients of the flux calculation, $\E{T}_2$
and $\E{\Th}_2$ (cf. eqs.
(\ref{trace-def}),(\ref{pressure-def}) below),
will be called ``basic components''
of the EM tensor.

It had been known for a long time \cite{chf77} that 2d minimally
coupled massless scalars -- i.e. in the absence of a dilaton
field -- provide the correct flux from the $2d$ conformal anomaly
$\E{T}_2$ alone. But an actual computation of the missing piece 
$\E{T^{\theta}{}_{\theta}}_2$ was not available until the
work of \cite{kuv98}.

The determination of $\E{T}_2$ precisely fits into
the elegant formalism of heat kernel regularisation\cite{gil75,aps76},
because in 2d ``by chance'' $\E{T}_2$ is simply the trace anomaly
of a massless scalar field which is related to the variation of the
effective action for a multiplicative (conformal!) factor of the
Laplace operator. By contrast,
the quantum correction $\E{T^{\theta}{}_{\theta}}_2$ cannot be computed
as easily in this formalism, because it lacks that key property. For this
reason in refs. \cite{kuv98,kuv99} the Laplace operator, including
a general coupling to the dilaton field, has been split into a product
of two Dirac operators whose combined determinant need be evaluated
in flat space only. Then a (multiplicative!) variation with respect to
the dilaton field is observed and allows the computation of the
component $\E{T^{\theta}{}_{\theta}}_2$,
dubbed ``dilaton anomaly'' in that work. Using both basic components
in a (``dilaton-deformed'') EM conservation equation \cite{baf98,lmr99}
produced exactly the black body Hawking flux at infinity from 2d
minimally coupled scalars which would follow from
the Hawking temperature computed from, say, the surface gravity
at the horizon. The outgoing flux
at infinity precisely coincided with the Stefan Boltzmann law
for $d=2$, which is related easily to the equally correct 4d result.
The only further input had been the vanishing of the asymptotic
incoming flux together with the condition of finiteness (or at least
integrability) of the flux at the horizon (Unruh vacuum) in global
Kruskal-Szekeres coordinates. In addition, by direct functional integration
an expression for the total effective action in the presence of a
general dilaton coupling was obtained. Despite of these satisfactory
final results, the step splitting the Laplace operator into two
linear Dirac operators from a rigorous mathematical point of view
seemed to be a doubtful one.

Therefore, in our present paper we attempt
to close that loop hole in an alternative derivation by a new application of
the covariant perturbation theory, introduced by Barvinsky and Vilkovisky (BV)
\cite{bav87,bav90}. This technique allows to proceed directly to the
effective action. In ref. \cite{bgv95} it already has been shown that
the BV effective action reproduces the correct trace anomaly of
conformally coupled scalar fields in $d=4$.
Actually, it contains more information
than just the trace anomaly, although the latter in general is the only
expectation value that can be calculated directly. In particular, the
dilaton effective action derived in the present work not only
produces the correct 2d trace anomaly but also represents another
derivation of the dilaton anomaly. However, beside the expression found in
\cite{kuv98,kuv99} we encounter an important IR-renormalisation effect,
i.e. something which had been by-passed altogether in the only
UV-sensitive previous approach.

That action is used by us only to determine the basic
components $\E{T}_2$ and $\E{T^{\theta}{}_{\theta}}_2$,
while the remaining components are reconstructed by integration of
the EM conservation equation, following the procedure introduced
by Christensen and Fulling (CF) \cite{chf77}.

In Section 2 we present the dilaton model and some
characteristic features of spherically reduced (SR) gravity.

Section 3.1 is devoted to the computation of the effective action
of the dilaton model by the covariant perturbation theory of
refs. \cite{bav87,bav90}.

In Section 3.2 we discuss the ambiguity of the non-local
effective action by a Green function perturbation theory
and fix it by appropriate boundary conditions and some infrared
regularisation.

The expectation values of the basic components to be derived
from the effective action are the subject of Section 4 and the
remaining components
in the Unruh state are determined by fixing the constants
$Q$ and $K$ of ref. \cite{chf77} accordingly.

In the Conclusions (Section 5) we summarize and discuss the obtained
results.

This paper further contains four Appendices: in Appendix A we shortly
present the SR procedure and derive some useful formulas.
Appendix B contains the second and third order of the Green function
perturbation theory. In Appendix C we show the (non-)conservation of the
2d dilaton EM tensor at the quantum level. In Appendix D the
regularisation of the heat kernel is demonstrated.

Important basic calculations and concepts of this paper can be found in
more detail in the PhD thesis \cite{hof02} of one of the authors
(D. H.), however a more careful discussion of (UV \emph{and} IR)
renormalisation issues is presented here.

\section{Dilaton Model}

A massless scalar field $S$ is considered on a four-dimensional
Schwarzschild spacetime $M$ with coordinates $x^{\mu}=(x^{\alpha},\theta,\vp)$,
in 4d coupled minimally (but not conformally!) to gravity:
\begin{equation}
L=\int_M\left[\frac{c^2}{16\pi G}R^{(4)}+\frac{(\n S)^2}{2}\right]
\sqrt{-g^{(4)}}d^4x\label{scalar action}
\end{equation}
In the following we set $c=G=\hbar=k_B=1$. This model can be spherically
reduced to a dilaton model on a two-dimensional spacetime $L$
by integrating out the isometry coordinates $\theta,\vp$, (cf. Appendix A
(\ref{R-sph.red.4d})):
\begin{equation}
L_{dil}=\int_L\left\{XR+\frac{(\n X)^2}{2X}-2
+X\left[\frac{(\n S)^2}{2}\right]\right\}\sqrt{-g}d^2x
\label{2daction}.
\end{equation}
On a four-dimensional spherically symmetric spacetime the EM tensor
exhibits only four independent components \cite{chf77}
($T^2{}_2=\Th=T^{\vp}{}_{\vp}$) in a vierbein basis
\begin{equation}
T^m{}_n=\left(\begin{array}{cccc}
T^0{}_0 & T^0{}_1 & 0 & 0 \\
-T^0{}_1 & T^1{}_1 & 0 & 0 \\
0 & 0 & T^2{}_2 & 0 \\
0 & 0 & 0 & T^2{}_2 \end{array}\right)\label{s-wave-1},
\end{equation}
where the first block (up to a factor $4\pi X$) equals the
two-dimensional EM tensor $T^{\alpha}{}_{\beta}$ on $L$.
Diffeomorphism invariance in the
dilaton model (\ref{2daction}) on-shell for $S$ implies the 2d
``non-conservation equation''
\begin{equation}
\n_{\alpha}T^{\alpha}{}_{\beta}=-\frac{\n_{\beta}X}{\sqrt{-g}}
\frac{\delta L_{dil}}{\delta X},\label{non-cons-eq.}
\end{equation}
whose solution for the Schwarzschild metric $g_{tt}=-(g_{rr})^{-1}=\Sss$
\begin{eqnarray}
T^r{}_t&=&-\frac{K}{M^2}\label{2dsol1}\\
T^r{}_r&=&\frac{1}{\Sss}\biggm\{\frac{Q-K}{M^2}\nonumber\\
&&+\int_{2M}^{r}\left[\frac{M\cdot T}{(r')^2}-\left(1-\frac{2M}{r'}\right)
\frac{\p_rX}{\sqrt{-g}}\frac{\delta L_{dil}^m}{\delta X}\right]dr'\biggm\}
\label{2dsol2}
\end{eqnarray}
only depends on the integration constants $Q,K$ and the unknown
functions $T,\frac{\delta L_{dil}^m}{\delta X}$ to be identified with
the basic components. Eq. (\ref{non-cons-eq.}) is just another
expression for the 4d conservation equation \cite{chf77} if one identifies
\begin{equation}
\Th=-\frac{1}{4\pi\sqrt{-g}}\frac{\delta L_{dil}}{\delta X}.
\end{equation}
This relation can be checked easily for the action (\ref{2daction})
using the definition of the EM tensor $T_{\mu\nu}=\frac{2}{\sqrt{-g}}
\frac{\delta L}{\delta g^{\mu\nu}}$. In Appendix C we show
that in a fixed classical background the non-conservation
equation also holds at the level of (renormalized) expectation values,
\begin{equation}
\n^{\beta}\E{T_{\alpha\beta}}_{ren}=\frac{\n_{\alpha}X}{2}
\E{m^2S^2-(\n S)^2}=-\frac{\n_{\alpha}X}{\sqrt{-g}}
\frac{\delta W}{\delta X}\label{EM.cons.quant.dil},
\end{equation}
where $W$ is the generating functional of connected Green
functions for the dilaton theory which at the one-loop level
coincides with the effective action (when the propagators of the external
lines are amputated). At the quantum level
the basic components are calculated by variation of the effective
action\footnote{It should be emphasized
that $W$ is a two-dimensional action and expectation
values derived from it could differ from those calculated in $4d$.}:
\begin{eqnarray}
\E{T}_2&:=&\E{T}=g^{\alpha\beta}\frac{2}{\sqrt{-g}}\frac{\delta W}
{\delta g^{\alpha\beta}}\label{trace-def}\\
\E{\Th}_2&:=&\E{\Th}=-\frac{1}{\sqrt{-g}}\frac{\delta W}{\delta X}
\label{pressure-def}
\end{eqnarray}
In the following all quantities are two-dimensional if no dimension index
is attached.
$T^r{}_t$ is the flux component of the EM tensor and differs
from the $4d$ flux (like all components of the EM tensor) by a factor
$(4\pi X)^{-1}$ (the dilaton $X$ is hidden in the spacetime measure
$\sqrt{-g_4}=X\sqrt{-g_2}$).
The constants $Q,K$ remain to be fixed by the boundary conditions
of the fields and are thus related to the quantum state of
the system \cite{chf77}.

All physical states are characterized by the choice $Q=0$ which
is a necessary condition for the finiteness of the EM tensor at the horizon
in global coordinates. The Hartle-Hawking state $\St{H}$ is given by
vanishing total flux $K_H=0$ (thermal equilibrium), whereas the Unruh
state $\St{U}$ is determined by vanishing incoming flux (leading to a non-zero
$K_U$). The (unphysical) Boulware state $\St{B}$ is defined by vanishing fields
in the asymptotic region. This is accomplished by setting $K_B=0$
and fixing $Q_B\neq0$ appropriately. Although not to be interpreted as a
\emph{physical} state, because of its natural boundary
conditions, in a certain sense it nonetheless is the natural state
of the effective action. On the one hand, in order to represent a
well-defined integral over the fields, natural (vanishing) asymptotic
values for them are necessary. On the other hand, in the path integral
(cf. (\ref{gen.functional}) below) the field
$S'$ is a sum of a classical solution $S_0$ and a quantum correction
$S_q$. The standard procedure is to set $S_0=0$ -- otherwise one
would have surface terms that would make the application of the
heat kernel method very difficult. This means that the incoming and outgoing
states correspond to the vacuum, i.e. the Boulware state.
If the expectation value of the EM tensor is calculated from
the effective action with the Boulware state values of $K_B=0$ and $Q_B$,
any other quantum state with $K=K_B+\tilde{K}, Q=Q_B+\tilde{Q}$
can be reconstructed by simply adding (to the first block in
(\ref{s-wave-1})) a term
\begin{equation}
\E{\tilde{T}^{\mu}{}_{\nu}}=\large\frac{1}{M^2}\left(\begin{array}{cc}
\frac{\tilde{K}-\tilde{Q}}{\Sss} & \frac{\tilde{K}}{\Sss^2}\\
-\tilde{K} & \frac{\tilde{Q}-\tilde{K}}{\Sss}
\end{array}\right)\normalsize\label{EM-tensor,diff}
\end{equation}
which is a special solution of the (non-)conservation equation.
Of course, this procedure works only if the basic components are insensitive
with respect to the state of the effective action which is true, as
long as the radiation does not affect the spacetime geometry
significantly, i.e. in the quasi-static phase of a BH \cite{hof02}.

\section{Non-Local Effective Action}

\subsection{Covariant Perturbation Theory}

The relation \cite{gil75,aps76} between the Euklidean effective
action and the heat kernel
$e^{-{\cal O}\tau}$ for the differential operator $\cal O$ is
given by\footnote{We denote Euklidean objects by an index $\cal E$.}
\begin{equation}
W_{\cal E}[g]=-\frac{1}{2}\frac{d\zeta[s]}{ds}\biggm|_{s=0}=
-\frac{1}{2}\frac{d}{ds}\frac{1}{\Gamma(s)}
\int_0^{\infty}\frac{d\tau}{\tau^{1-s}}
\text{tr}\,e^{-{\cal O}\tau}\biggm|_{s=0}.\label{zeta-function}
\end{equation}
The trace of the heat kernel may be expressed in a coordinate
basis
\begin{equation}
\text{tr}\,e^{-{\cal O}\tau}=\int_M\left<x\right|e^{-{\cal O}\tau}
\left|x\right>\sqrt{g}d^4x=\int_M G_{\cal O}(x,x;\tau)\sqrt{g}d^4x.
\end{equation}
In the most common applications the heat kernel is expanded around
$\tau=0$ \cite{gil75,aps76}.
In contrast to that, the aim of the method developed in
\cite{bav87,bav90,bgv95} is to use (\ref{zeta-function}) directly in
order to find an
expression of the heat kernel which is valid for all values
of $\tau$. This allows performing the $\tau$-integration
and computing the effective action for any Euklidean Laplacian
\begin{equation}
{\cal O}=-\triangle-E,
\end{equation}
where $\triangle=g^{\mu\nu}\n_{\mu}\n_{\nu}$ is the contraction
of two general covariant derivatives (which may include a gauge part)
by some Euklidean metric $g$, and $E$ is an endomorphism, i.e. some
linear bounded map from the space of fields into itself.
The covariant perturbation series is based on a separation
of the spacetime metric into a flat part $\tilde{g}$ and a
perturbing part $h$: $g_{\mu\nu}=\tilde{g}_{\mu\nu}+h_{\mu\nu}$
(we use the notation of BV). Nevertheless, each order can
be represented by covariant expressions corresponding to the
full metric $g$, such as the scalar curvature $R$, because the
flat metric does not produce
gravitational effects $R(\tilde{g})=0$. Then one expands the Laplacian
and the heat kernel in orders of $h$:
\begin{eqnarray}
{\cal O}&=&-\triangle_0+h^{\mu\nu}\tilde{\n}_{\mu}\tilde{\n}_{\nu}+\dots,\\
G_{\cal O}(\tau)&=&e^{-{\cal O}\tau}
=\sum_{n=0}^{\infty}G^n_{\cal O}(\tau),\label{cov.pert.series}
\end{eqnarray}
where $G^0_{\cal O}(\tau)=e^{\tau\triangle_0}$ and $\triangle_0$
is the flat Laplacian. Up to the second order in the curvature
the trace of the heat kernel in even dimensions
$d=2\omega, \omega\in\mathbb{N}$ is found to be\footnote{In the
following we set $\omega=1$, i.e. $d=2$.} \cite{bav90}
\begin{multline}
\text{tr}\,e^{-{\cal O}\tau}=\frac{1}{(4\pi\tau)^{\omega}}
\int_M\text{tr}\biggm\{\1+\tau\left(\frac{R}{6}+E\right)\\
+\tau^2\biggm[R\left(\frac{1}{16(-\tau\triangle)}
+\frac{f(-\tau\triangle)}{32}
+\frac{f(-\tau\triangle)-1}{8(-\tau\triangle)}
+\frac{3[f(-\tau\triangle)-1]}{8(\tau\triangle)^2}\right)R\\
+E\left(\frac{f(-\tau\triangle)}{6}+\frac{f(-\tau\triangle)-1}
{2(-\tau\triangle)}\right)R+R\frac{f(-\tau\triangle)}{12}E
+E\frac{f(-\tau\triangle)}{2}E\biggm]\biggm\}\sqrt{g}d^{2\omega}x,
\label{nonlocal.effective.action}
\end{multline}
where
\begin{equation}
f(x)=\int_0^1e^{-a(1-a)x}da.\label{f2}
\end{equation}
In the present paper we restrict ourselves to the second
order of covariant perturbation theory, i.e. terms up to $R^2,ER,E^2$.
This is at least sufficient
to compute the \emph{exact} trace anomaly which is completely determined
by a single term of the local Seeley-DeWitt expansion \cite{bgv95,aps76}
corresponding to that order. With respect to other expectation values
like the dilaton anomaly the necessity for higher orders cannot
be excluded, a priori, though.

For $d=2$ the trace of the heat kernel
(\ref{nonlocal.effective.action}) (to this order) in
(\ref{zeta-function}) produces five types of integrals.
They contain IR or (and) UV divergences that have to be
regularized by restricting the range of the $\tau$-integration as
$\int_{\ve}^T d\tau$ in the limit $T\to\infty,\ve\to0+$.
Examples for these rather tedious calculations are shown in
Appendix D. They can be done analytically to leading order in $T,\ve$
for these IR, resp. UV regularisation parameters, and for the
corresponding next finite terms.
The formal contribution of these terms to the effective action
(\ref{zeta-function}) is given by
\begin{eqnarray}
\frac{d}{ds}\left\{\frac{1}{\Gamma(s)}\int_0^{\infty}\tau^{s-1}
d\tau\right\}\biggm|_{s=0}&=&\ln\frac{T}{\ve}\hspace{1.9cm}IR,UV\label{div1}\\
\frac{d}{ds}\left\{\frac{1}{\Gamma(s)}\int_0^{\infty}\tau^{s-2}
d\tau\right\}\biggm|_{s=0}&=&\frac{1}{\ve}\hspace{3.1cm}UV\label{div2}\\
\frac{d}{ds}\left\{\frac{1}{\Gamma(s)}\int_0^{\infty}\tau^{s}
f(-\tau\triangle)d\tau\right\}\biggm|_{s=0}&=&
\frac{2\cdot\left[\ln(-T\triangle)+\gamma_E\right]}{-\triangle}
\hspace{0cm}IR\label{div3}\\
\frac{d}{ds}\left\{\frac{1}{\Gamma(s)}\int_0^{\infty}\tau^{s-1}
\frac{f(-\tau\triangle)-1}{-\triangle}d\tau\right\}\biggm|_{s=0}&=&
\frac{2-\ln(-T\triangle)-\gamma_E}{-\triangle}\hspace{0cm}IR\label{div4}\\
\frac{d}{ds}\left\{\frac{1}{\Gamma(s)}\int_0^{\infty}\tau^{s-2}
\frac{f(-\tau\triangle)-1}{\triangle^2}d\tau\right\}\biggm|_{s=0}&=&
\frac{\ln(-T\triangle)+\gamma_E-\frac{8}{3}}{6(-\triangle)}
\hspace{0cm}UV\label{div5}
\end{eqnarray}
where $\gamma_E\approx0,57721$ is the Euler constant. Collecting all
terms in the heat kernel (\ref{nonlocal.effective.action}) it turns
out that the contributions from the divergent parts $\ve^{-1},\ln\ve$,
and $\ln T$ surprisingly \emph{cancel to zero} in those nonlocal terms
containing the scalar curvature $R$. Thus, the most general
two-dimensional regularized effective action to second order of
perturbation theory in $R$, resp. $E$ reads
\begin{equation}
W_{\cal E}^{reg}[g]=\frac{1}{96\pi}\int_L\biggm[12c_0-c_1(2R+12E)
+(R+12E)\frac{1}{\triangle}R
+12E\frac{c_2}{\triangle}E\biggm]\sqrt{g}d^2x.\label{non-local.eff.action}
\end{equation}
Here we have introduced the following regularisation terms:
$c_0=\ve^{-1},c_1=\ln(\ve/T),c_2=\ln(-T\triangle)+\gamma_E$.
Actually the term $\propto c_1R$ can be dropped, being a total
divergence. Note that $c_2$ beside a logarithmic divergence contains
an ill-defined expression in the Laplacian. In Section 4 the (eventual)
contributions of the regular and divergent terms will be discussed separately.

\subsection{Effective Dilaton Action}

So far all steps in this section were valid for general Laplacian
$\triangle$ and endomorphism $E$.
In order to establish the effective action of the dilaton model
(\ref{2daction}) we have to specify $\triangle$ and $E$ accordingly,
while returning to Lorentzian spacetime.
To achieve this we must reconsider the four-dimensional generating
functional, determined by the matter part of (\ref{scalar action})
\begin{equation}
Z[g_4]={\cal N}\int{\cal D}\left(\sqrt[4]{-g_4}S\right)\cdot e^{iL_m^4[g_4,S]}
={\cal N}\int{\cal D}\left(\sqrt[4]{-g_4}S\right)
\cdot e^{-i\int_M S\square S\sqrt{-g_4}d^4x}.
\end{equation}
${\cal N}$ is a normalization constant and the factor $\sqrt[4]{-g}$
in the path integral measure establishes diffeomorphism invariance
\cite{fln88} of the path integral. The SR generating functional is
obtained by introducing the SR d'Alembertian $\square_4=\square_2
+\frac{\n X}{X}\n$ (cf. (\ref{Laplace-sph.red.}) in the Appendix for $d=4$)
and measure $\sqrt{-g_4}=X\sqrt{-g_2}$ in the classical action
and by integration over the angular coordinates $\theta,\vp$:
\begin{equation}
Z[g_2]={\cal N}\int{\cal D}\left(\sqrt[4]{-g_2}\sqrt{X}S\right)
\cdot e^{-4\pi i\int_LS(X\square_2+\n X\n)S\sqrt{-g_2}d^2x}.
\label{gen.functional}
\end{equation}
It is now convenient to define a new field $S':=\sqrt{X}S$ such
that\footnote{In \cite{hof02}
the original field $S$ was preserved, leading to an additional
dilaton factor $X$ from the measure
during spherical reduction. In the 2d action this difference can be
described by a conformal transformation of the metric by this factor,
not affecting the Hawking flux but other components of the EM tensor.}
$L_m[g_2,S']=-\int_LS'{\cal O}_{\cal M}S'\sqrt{-g_2}d^2x$ defining
the complete d'Alembertian of the dilaton model\footnote{$\cal M$
indicates Minkowski signature.}
\begin{eqnarray}
{\cal O}_{\cal M}&=&\frac{1}{\sqrt{X}}(X\square_2+\n X \n)\frac{1}{\sqrt{X}}
=\square_2+E_{\cal M},\label{O tot}\\
E_{\cal M}&=&\frac{(\n X)^2}{4X^2}-\frac{\square X}{2X}.
\end{eqnarray}
Inserting these results into (\ref{non-local.eff.action})
and going back to Lorentzian spacetime
$d\tau=idt,W_{\cal M}=iW_{\cal E},\triangle\to-\square,
R_{\cal E}\to-R_{\cal M},E_{\cal E}\to-E_{\cal M}$
the effective action of the dilaton model (\ref{2daction})
follows\footnote{In the following we omit the index $dil$.}:
\begin{multline}
W_{\cal M}^{dil}[g]=\frac{1}{96\pi}\int_L\Biggm\{-12c_0-
3c_1\left[\left(\frac{\n X}{X}\right)^2-2\frac{\square X}{X}\right]\\
+\left[R+3\left(\frac{\n X}{X}\right)^2-6\frac{\square X}{X}\right]
\frac{1}{\square}R\\
+\frac{3}{4}\left[\left(\frac{\n X}{X}\right)^2-2\frac{\square
X}{X}\right]\frac{c_2}{\square}\left[\left(\frac{\n X}{X}\right)^2
-2\frac{\square X}{X}\right]\Biggm\}\sqrt{-g}d^2x.\label{dilaton.eff.action}
\end{multline}
A particular, attractive feature of the 2d dilaton model is that
most of its effective action (\ref{dilaton.eff.action}) can be brought
into a local form by choosing a conformal
gauge $g_{\alpha\beta}=e^{2\rho}\eta_{\alpha\beta}$ of the spacetime metric.
The scalar curvature then becomes $R=-2\square\rho$ and the endomorphism
$E=\square\phi-(\n\phi)^2$. It is convenient
to represent the dilaton field in the form $X=e^{-2\phi}$.
If one (naively) uses the relation
\begin{equation}
\square^{-1}\square=1\label{nonloc.rel.}
\end{equation}
one obtains the local part of the effective dilaton action
\begin{equation}
W_{l}[g]=\frac{1}{24\pi}\int_L\Bigm\{-3c_0+3c_1(\n\phi)^2
+\rho\square\rho+6\rho(\n\phi)^2-6\rho\square\phi\Bigm\}\sqrt{-g}d^2x
\label{loc.dilaton.eff.action}
\end{equation}
which turns out to be identical to the one derived in
\cite{kuv98}. However a new contribution
\begin{equation}
W_{nl}[g]=\frac{1}{24\pi}\int_L\Bigm\{3\left[\square\phi-
(\n\phi)^2\right]\frac{c_2}{\square}\left[\square\phi-
(\n\phi)^2\right]\Bigm\}\sqrt{-g}d^2x
\label{nonloc.dilaton.eff.action}
\end{equation}
appears in our present approach which cannot be brought into a local
form. Nevertheless, we will be able to show below that the relevant
expectation value derived from (\ref{nonloc.dilaton.eff.action}) is
local after all but ill-defined. It is remarkable that the
divergent terms do not contain the scalar curvature.

Obviously, the form (\ref{loc.dilaton.eff.action}) is not unique.
Namely, by naively using relation (\ref{nonloc.rel.}) one implicitly
disregards a homogeneous solution $\chi$ of the wave equation
$\square\chi=0$. It is the aim of the next
section to present a heuristic argument that the proper choice is indeed
$\chi=0$, and thus (\ref{nonloc.rel.})
is in agreement with the boundary conditions which we should impose
onto the scalar field $S$.

\subsection{Homogeneous Solution and Boundary Conditions}

The relation of a particular choice of $\chi$
to the boundary conditions of the (massless) scalar field $S$
can be seen when writing the inverse d'Alembertian as an
integral over the Green function of $S$ and
applying Green's theorem whereby $\square G(x,x')=-\delta(x-x')$ and
$\square f=F$:
\begin{multline}
-\frac{1}{\square}F=\int_LG(x,x')F(x')\sqrt{-g'}d^2x'\\
=-f(x)-\oint_{\p L}\Bigm[f\n_{\alpha}'G-G\n_{\alpha}'
f\Bigm]\sqrt{-g'}\ve^{\alpha}{}_{\beta}(dx')^{\beta}\\
=-f(x)-\int_{2M}^{\infty}f(x')\p_{t'}G\frac{dr'}
{\left(1-\frac{2M}{r'}\right)}\biggm|_{t'=\infty}
+\,{\dots}\biggm|_{t'=-\infty}\\
-\int_{-\infty}^{\infty}\Big[f(x')\p_{r'}G-G\p_{r'}f(x')\Big]
\left(1-\frac{2M}{r'}\right)dt'\biggm|_{r'=2M}
+\,{\dots}\biggm|_{r'=\infty}.\label{bound.basic.int}
\end{multline}
In the step to the third line the explicit form of the Schwarzschild
metric has been used. According to (\ref{bound.basic.int}),
to guarantee regularity of the boundary terms, $G(x,x')$
must vanish at least linearly for $r'=2M$.
This is a natural condition for the Green functions
since the manifold $L$ is in fact a half-plane that is bounded by the
coordinate singularity $r=2M$.
Further, one must employ some infrared regularisation to
render finite the support of the Green functions.
As the flux is measured at some finite distance $r$ from
the BH at some instant of time $t$ during the quasi-static phase
all boundary terms should vanish after that infrared regularisation.
This can be realized by simply \emph{dropping all boundary
terms}\footnote{Alternatively one could introduce a finite mass parameter or
consider a manifold of finite size.}. But then (\ref{bound.basic.int})
reduces to $-\frac{1}{\square}F=-f$ (provided that $f$ has
at most logarithmic divergences on the horizon) and relation
(\ref{nonloc.rel.}) is fulfilled.

It is instructive to analyze expressions like
$\square^{-1}$ more explicitly. Unfortunately, the Green functions
on a Schwarzschild spacetime cannot be obtained in a closed form.
Nevertheless, one may construct a heuristic argument by considering
the properties of a formal perturbation series
\begin{multline}
G(x,x')=G_0(x,x')+\int''_LG_0(x,x'')\delta\square''G_0(x'',x')d^2x''\\
+\int''_L\int'''_LG_0(x,x'')\delta\square''G_0(x'',x''')
\delta\square'''G_0(x''',x')d^2x''d^2x'''+\dots,\label{pert.Green}
\end{multline}
where
\begin{equation}
\square=\Sss^{-1}\p_t^2-\p_r\left[\Sss\p_r\right]
\end{equation}
is the two-dimensional Schwarzschild d'Alembertian and
\begin{equation}
\delta\square=\square-\square_0=\frac{2M}{r-2M}\p_t^2+\p_r\left(\frac{2M}{r}
\p_r\right)\label{pert.Laplace}
\end{equation}
is the difference to $\square_0$, the flat one.

In eqs. (\ref{pert.sec.order}), resp. (\ref{pert.thi.order}) of
Appendix B the results for the second, resp.
the third order of the perturbation series are given.

The Green functions on the flat half-plane which fulfill the
imposed boundary conditions are easy to construct, singling out
the appropriate eigenfunctions. For instance, the flat retarded
Green function reads
\begin{eqnarray}
G^{ret}_0&=&\theta(t-t')G_0^{(0)}\label{G-ret}\\
G_0^{(0)}&=&\frac{1}{2}[\theta(r-r'+t'-t)-\theta(r-r'+t-t')\nonumber\\
&&-\theta(r+r'-4M+t'-t)+\theta(r+r'-4M+t-t')].
\end{eqnarray}
In the same manner the flat advanced and Feynman-type ``causal'' Green
functions can be obtained. It should be noted that the non-local terms
in the effective action, having the form $\int_LG(x,x')F(x')\sqrt{-g}d^2x'$
with \emph{time-independent} $F=F(r)$,
are \emph{independent of the type of Green function}
as can be verified easily. Thus only the retarded Green function (\ref{G-ret})
need to be considered and any further ambiguity of that type disappears.

Among the non-local expressions that appear in the
effective action (\ref{dilaton.eff.action}) the first one
has the form $\square^{-1}R$. The first three orders
of the perturbation series with (\ref{pert.Green}) and the formulas
(\ref{pert.sec.order}),(\ref{pert.thi.order}) of Appendix B yield
\begin{multline}
\int_{L}G(x,x')R(x')\sqrt{-g'}d^2x'=1+\frac{1}{2}
+\frac{1}{3}+\dots-\frac{2M}{r}-\frac{2M^2}{r^2}
-\frac{8M^3}{3r^3}-\dots\\
\to\ln\Sss-\ln0=2\rho-\ln\tilde{\ve},\label{basic-int.1}
\end{multline}
$\tilde{\ve}\to0$, which is suggested by the formal
summation in the first line.
Bearing in mind that $R=-2\square\rho=-\square\ln\Sss$ we may argue
that the perturbation series produces a homogeneous solution
which is just an infinite constant $\chi_{\rho}=-\frac{1}{2}\ln\tilde{\ve}$.
Its appearance could have been expected also as it is ``required''
to shift the absolute value of the integral (\ref{basic-int.1}) in
accordance with the boundary conditions: if one inserts $r=2M$ on the l.h.s.
the value of the integral must become zero. The action of the
inverse d'Alembertian on the half-plane on a function $F=\square f$
thus becomes
\begin{equation}
\square^{-1}\square f(r)=f(r)-f(2M).\label{nonloc.rel.final}
\end{equation}
But this is consistent with the proper definition of a delta-function on
the half-plane:
\begin{equation}
\delta^{hp}(x-x'):=\delta(r-r')\delta(t-t')-\delta(r+r'-4M)\delta(t-t')
\label{delta-hp.}
\end{equation}
Using $\delta^{hp}$ in Green's theorem
(\ref{bound.basic.int}) immediately yields indeed
(\ref{nonloc.rel.final}). In any case this boundary term does not change the
effective action (\ref{loc.dilaton.eff.action}) because all homogeneous
\emph{constants} can be absorbed anyway by the renormalisation
constant $c_1$, which, on the other hand, has no effect on the Hawking
flux (c.f. (\ref{flux-ind.}) below).

Finally, the first order of the second type of non-local
term $\square^{-1}\square\phi=\square^{-1}\left(\frac{4M-r}{r^3}\right)$
in the effective action should be checked:
\begin{equation}
\int_{L}G_0(x,x')\frac{4M-r'}{(r')^3}d^2x'
\approx\ln\left(\frac{r}{2M}\right)-\Sss+...
\end{equation}
The leading order $\ln\left(\frac{r}{2M}\right)$
corresponds to the expected result as $\phi=-\ln r$. To this order
nothing can be said about the additional
term $-\Sss$ which should be absorbed by
higher orders of the perturbation series.

\section{Hawking Flux}

\subsection{Regular Part}

Having derived in Section 2 the effective action of the
dilaton model after fixing its local form
(\ref{loc.dilaton.eff.action}) by considering boundary
conditions, it is now straightforward
to calculate the expectation values of the basic components by functional
variation for $\rho$ and $X$ (or, more conveniently, $\phi$).
First we only regard the regular part $\propto\int_L[\rho\square\rho
+6\rho(\n\phi^2)-6\rho\square\phi]\sqrt{-g}d^2x$ of
(\ref{loc.dilaton.eff.action}).
The trace (\ref{trace-def}) then becomes
\begin{equation}
\E{T}=-\frac{1}{\sqrt{-g}}\frac{\delta W_{reg}}
{\delta\rho}=\frac{M}{3\pi r^3},\label{trace}
\end{equation}
and the pressure component (\ref{pressure-def}), remembering
that $X=e^{-2\phi}$,
\begin{equation}
\E{\Th}=\frac{1}{2X\sqrt{-g}}\frac{\delta W_{reg}}{\delta\phi}
=\frac{-1}{8\pi r^5}\left\{4M+(4M-r)\left[\ln\Sss-\ln\tilde{\ve}\right]
\right\}.
\label{pressure}
\end{equation}
The appearance of an infinite constant $\ln\tilde{\ve}$
can be traced back to the
particular boundary conditions used in this work to define the inverse
d'Alembertian (\ref{nonloc.rel.final}). It can be shifted to the (also
infinite) regularisation constant $c_1$, see next
Section\footnote{For the moment we keep this notation until we
consider the contribution of the corresponding divergent term in the
effective action.}.

As discussed in the Introduction, the basic components can be considered
as being state-independent, whereas the proper quantum state for a
well-defined effective action has been argued to be the static
(unphysical) Boulware state. Thus we must rely on the conservation
equation (\ref{EM.cons.quant.dil}) to compute the remaining
components of the EM tensor in the Unruh state. A necessary
condition for regularity of the EM tensor at the horizon
was $Q_U=0$. The other constant $K_U$
is determined by the condition that the incoming flux vanishes
\begin{multline}
\U{T_{++}}=\frac{\Sss}{4}\left(\E{T^t{}_t}-\E{T^{r_{\ast}}{}_{r_{\ast}}}
-2\E{T^{r_{\ast}}{}_{t}}\right)\\
=\frac{\Sss}{4}\left(\E{T}-2\E{T^r{}_r}-2\E{T^{r_{\ast}}{}_{t}}\right)\\
\stackrel{r\to\infty}{\to}\frac{K_U}{M^2}
-\frac{\int_{2M}^{\infty}\left[\frac{M}{(r')^2}\E{T}
+\left(1-\frac{2M}{r'}\right)\p_rX\E{\Th}\right]dr'}{2}-\frac{Q_U}{2M^2}=0.\\
\end{multline}
Inserting (\ref{trace}),(\ref{pressure}) and fixing the dilaton field
to its standard form $X=r^2$ in spherical coordinates we obtain
\begin{equation}
K_U=M^2\int_{2M}^{\infty}\left[\frac{M}{2(r')^2}\E{T}_2
+(r'-2M)\E{\Th}_2\right]dr'=-\frac{1}{768\pi}.
\end{equation}
By the CF equations (\ref{2dsol1}),(\ref{2dsol2}) the total
flux $F_{reg}$ from the regular part of the effective action through a
large spherical shell surrounding the BH is then given by
\begin{equation}
F_{reg}=\U{T^r{}_t}_2=-\frac{K_U}{M^2}=\frac{1}{768\pi M^2},
\label{Hawking.flux.tot}
\end{equation}
which precisely coincides with the result obtained in refs.
\cite{kuv98,kuv99}. The related flux in 4d through
a sphere of size $4\pi r^2$ becomes
\begin{equation}
\U{T^r{}_t}_4=\frac{1}{3072\pi^2M^2r^2}.\label{Hawking.flux.massless}
\end{equation}
As a consequence the BH behaves as a black body at Hawking temperature
with the radiation flux according to Stefan-Boltzmann's law
\cite{kuv98,kuv99}. It is appropriate,
though, to emphasize at this point the drawback of this solution to
the energy flow problem
\cite{baf99}: at the horizon the energy density and other components of the
EM tensor exhibit a logarithmic singularity in global coordinates!
As argued in \cite{kuv99} this singularity (being integrable) is not in
contradiction with the finiteness of the total flux. Actually, we also
obtain the outgoing flux in light-cone coordinates $T_{--}$ with that
singularity (eq. (97) of \cite{kuv99}) in our present approach
\begin{equation}
\U{T_{--}}=\frac{\Sss^2}{768\pi M^2r^2}\left\{r^2+4Mr+12M^2
+48\left[\ln\Sss-\ln\tilde{\ve}\right]\right\}.
\end{equation}

\subsection{Divergent Terms}

The first divergent term $\propto c_0=\ve^{-1}$ in
(\ref{loc.dilaton.eff.action}) is a pure UV divergence and has the form
of a cosmological constant:
\begin{equation}
W_{c_0}=\frac{-1}{8\pi}\int_Lc_0\sqrt{-g}d^2x
=\frac{-c_0}{8\pi}\int_Le^{2\rho}d^2x
\end{equation}
It contributes to the trace of the EM tensor
$\E{T}_{c_0}=\frac{c_0}{4\pi}$ and hence also to the asymptotic flux
($\ve$ has dimension $M^2$)
\begin{equation}
K_U^{c_0}=M^2\int_{2M}^{\infty}\frac{M\E{T}_{c_0}}{2r^2}dr
=\frac{M^2c_0}{16\pi}.
\end{equation}
This divergent contribution to the EM tensor can be interpreted
as an infinite vacuum energy because it even appears in the case of
flat spacetime $M=R=0$. A renormalized EM tensor can thus be defined
by subtracting the flat spacetime value (with flat metric $\eta$)
$\E{T_{\alpha\beta}}_{ren}:=\E{T_{\alpha\beta}}_{g_L}
-\E{T_{\alpha\beta}}_{\eta}$ or by simply setting $c_0=0$.

Further, we had an IR-UV divergence $\propto c_1=\ln(\ve/T)$
in (\ref{loc.dilaton.eff.action}), contributing only to the pressure component
\begin{equation}
\E{\Th}_{c_1}=\frac{c_1(r-4M)}{8\pi r^5}.
\end{equation}
Although the asymptotic behavior of the EM tensor is unaffected by $c_1$
because
\begin{equation}
K_U^{c_1}=M^2\int_{2M}^{\infty}(r-2M)\E{\Th}_{c_1}dr=0\label{flux-ind.}
\end{equation}
it produces infinite contributions to the EM tensor in higher orders in $r$.
Comparing with (\ref{pressure}) we observe that $c_1$ appears in the same
place as the homogeneous solution $\chi_{\rho}=-\frac{1}{2}\ln\tilde{\ve}$
needed to fulfill the boundary conditions (cf. Section 3.3) by
(\ref{nonloc.rel.final}). After all, any \emph{constant homogeneous solution
$\chi_{\rho}$ leaves the asymptotic flux invariant} and can be shifted
to the regularisation constant $c_1$.

The situation clearly is different if the homogeneous
solution is a \emph{function}. If, for instance, one chooses as in ref.
\cite{baf99}
\begin{equation}
\tilde{\chi}_{\rho}=\frac{1}{2}\left[\frac{r}{2M}-1
+\ln\left(\frac{r}{2M}-1\right)-\ln\tilde{\ve}\right],\label{chi-nontriv.}
\end{equation}
in order to eliminate the logarithmic singularity of the flux
at the horizon, this
would mean that the corresponding non-local term in
the effective action had the form (cf. (\ref{basic-int.1}))
\begin{equation}
\int_{L}G(x,x')R(r')d^2x'=\frac{r}{2M}-1
+\ln\left[\biggm(\frac{r}{2M}-1\biggm)
\biggm(1-\frac{2M}{r}\biggm)\right]-2\ln\tilde{\ve},
\end{equation}
and the asymptotic Hawking flux not only is affected, but even
would become \emph{negative} because
$\tilde{K}_U>0$:
\begin{equation}
\tilde{K}_U=K_U+M^2\int_{2M}^{\infty}\frac{(r'-2M)(r'-4M)}{4\pi(r')^5}
\tilde{\chi}_{\rho}{dr'}=K_U+\frac{1}{128\pi}=\frac{5}{768\pi}.
\label{BalbFabb}
\end{equation}
This is the result obtained in ref. \cite{baf99}, eq. (16). Also the
other components of the EM tensor calculated with $\tilde{\chi}_{\rho}$
in (\ref{chi-nontriv.}) can be verified to agree with those of that
work. As observed by the authors of \cite{baf99} themselves,
then the weak energy condition is clearly violated in the
asymptotic region. We do not believe that this serious consequence
of the choice (\ref{chi-nontriv.}) justifies its aim to eliminate
the logarithmic singularity at the horizon \cite{baf99}.

Finally, we have a non-local divergent part (\ref{nonloc.dilaton.eff.action})
of the effective action $\propto c_2$ whose contribution to the
basic components can be localized in a conformal gauge
\begin{eqnarray}
\E{T}_{c_2}&=&-\frac{1}{\sqrt{-g}}\frac{\delta W_{nl}}{\delta\rho}
=-\frac{1}{8\pi\sqrt{-g}}\frac{\delta}{\delta\rho}\int_LE
\frac{\ln(-Te^{-2\rho}\square_0)}{\square}E\sqrt{-g}d^2x\nonumber\\
&=&\frac{1}{4\pi}E\frac{1}{\square}E=\frac{\square
\rho[\rho-\rho(2M)]}{4\pi}=\frac{M[\ln\Sss-\ln\tilde{\ve}]}{4\pi r^3}\\
\E{\Th}_{c_2}&=&\frac{1}{2X\sqrt{-g}}\frac{\delta W_{nl}}{\delta\phi}
=\frac{1}{8\pi r^2\sqrt{-g}}\int_L\frac{\delta E}{\delta\phi}
\frac{c_2}{\square}E\sqrt{-g}d^2x\label{pressure-c2}\\
&=&\frac{[\square+2\square\phi+2\n\phi\n]}{8\pi r^2}\frac{c_2}{\square}E
=\frac{[\square+2\square\phi+2\n\phi\n]c_2[\rho-\rho(2M)]}{8\pi r^2},
\nonumber
\end{eqnarray}
where we have used the relation (after having varied the effective action!)
\begin{equation}
E=\square\phi-(\nabla\phi)^2=\frac{4M-r}{r^3}-\frac{2M-r}{r^3}=\frac{2M}{r^3}
=-\frac{R}{2}=\square\rho.
\end{equation}
(\ref{pressure-c2}) contains an ill-defined
expression $\ln(T\square)$ that cannot be treated further. Even if $c_2$
were only a constant, the pressure component would change by a term
$\E{\Th}_{c_2}=\frac{c_2}{8\pi r^5}\{4M+(4M-r)[\ln\Sss
-\ln\tilde{\ve}]\}$ identical to the original expression (\ref{pressure}).
In that case the asymptotic flux would be affected too:
$F_{reg}\to F_{reg}\cdot(1-3c_2)$. Because $F_{reg}$ is supposed to be
the correct result this constant had to be zero.

The appearance of IR divergences within the covariant perturbation
theory (in contrast to the local Seeley-DeWitt expansion used in former
derivations \cite{kuv98,kuv99}) is not surprising as it allows for
infinitely large values of the eigentime $\tau$, corresponding to
zero modes. As long as a more detailed analysis of these
terms does not exist we can only highlight the existence of such terms
while assuming that their (infinite) contribution to the EM tensor
might be canceled by some mechanism that has not been considered yet
and may necessitate the inclusion of (even arbitrary ?) higher orders in
covariant perturbation theory.

\section{Conclusions and Outlook}

The Hawking flux from a spherical
Black Hole has been reconsidered, whereby we followed the line of
solving the EM conservation law, as proposed by Christensen
and Fulling many years ago \cite{chf77}. We have linked the formalism
of $d=4$ to the ``non-conservation equation'' in the effective $d=2$
dilaton theory where as in $4d$ gravity, beside two constants,
only two ``basic components'' are a necessary input, the 2d trace
anomaly $\E{T}_2$ of the EM tensor and the
``dilaton anomaly'' $\E{\Th}_2$ to be interpreted also as (part of)
the pressure component
$\E{\Th}_4=\E{T^{\vp}{}_{\vp}}_4$ in $d=4$.

Whereas the computation of $\E{T}_2$
is known for a long time to fit perfectly into the heat kernel formalism,
in its single previous determination of refs. \cite{kuv98,kuv99}
$\E{\Th}_2$ only by a tour de force argument had been made
accessible to that technique.
In our present work we replace \emph{both}
derivations by a new application of the ``covariant perturbation
theory'', introduced by Barvinsky and Vilkovisky \cite{bav87,bav90},
which allows the direct determination of the effective action also
in the presence of dilaton fields. Again the heat kernel is used, albeit
in a slightly different manner. The non-local form of this action
implies an important dependence on boundary conditions for the Green
functions of the scalar field in the given background. We argue
that the effective action only has a consistent interpretation when
\emph{for that quantity} the (otherwise unphysical) Boulware state
determines the boundary condition at infinity (infrared regularisation).
This is supported by a study of the Green functions where -- in a formal
perturbation series starting from flat spacetime -- a natural boundary
condition at the horizon directly leads to a \emph{constant homogeneous
solution}. The latter can be absorbed in a renormalisation constant
which does not affect the asymptotic flux. Generically, the choice
of different homogeneous solutions represents
the ambiguity inherent in the non-local effective action.

From the regular part of that effective action in the Boulware
state the Hawking
flux is \emph{not} derived directly, but only the non-radiative
basic components $\E{T}_2$ and $\E{\Th}_2$ which are, indeed,
\emph{independent} of the assumed (Boulware, Hartle-Hawking,
Unruh) quantum state  \cite{hof02}. Inserting them into the
EM (non-)conservation equation the correct Hawking flux in the
Unruh state (in agreement with the black body derivation)
is obtained, without proceeding through one mathematically
questionable step which in refs. \cite{kuv98,kuv99} produced the
same physically reasonable result. 

We also clarify the consequences of the ambiguity brought
about by different \emph{other} choices of that homogeneous solution
which are not constants.
They may even yield a negative flux \cite{baf99}.
A choice like the one in that work, as noted by the authors themselves,
violates the weak energy condition which we consider a more serious defect
than the (logarithmic, hence integrable) divergence of some of the
components of the EM tensor at the horizon \cite{kuv98,kuv99}.

Beside the above-mentioned attractive features of the covariant
perturbation theory, it reveals the existence of three new divergent terms
(up to this order in curvature) which require further investigation.
The first one $\propto c_0$ is interpreted as an infinite vacuum energy
and its contribution to the flux can be removed by common renormalisation
arguments. The second one $\propto c_1$ seems to be related to the
boundary conditions as it contributes another logarithmic (UV)
divergence at the horizon, as compared to the one produced by the
homogeneous solution $\chi_{\rho}$, leaving
a logarithmic IR divergence. Although we could show the independence
(\ref{flux-ind.}) of the asymptotic flux of $c_1$ as well as of any
\emph{constant} $\chi_{\rho}$, the presence of an infinite constant in
higher order terms in $r$ could not be avoided (at least, when renormalizing
this constant to zero one is left with the logarithmic divergence
at the horizon as in \cite{kuv98,kuv99}). Finally, our approach
yields a completely new term (\ref{nonloc.dilaton.eff.action})
for the effective action that cannot be localized even in conformal
gauge and contains an IR divergence $c_2$ which is coupled to an
expression $\ln\square$. Although it could be shown that its
contribution to the EM tensor \emph{can} be given in a local form, the latter
could not be evaluated further due to its ill-definedness.
To sum it up, the covariant perturbation theory (by its non-local
character) produced new IR divergent terms that were absent in the
local Seeley-DeWitt expansion used in former derivations
\cite{kuv98,kuv99}. A proper renormalisation of these terms would
require a detailed analysis of this method (e.g. considering higher orders
in curvature or partial summations) which might be the issue of future work.

Beyond the application to the present problem the extension of covariant
perturbation theory may set an example for the use of that
technique in theories where matter is also coupled non-minimally to
a scalar field (such as scalar-tensor theories). 
Also applications to a non-perturbative approach,
where the geometry is integrated out, seem promising \cite{gru01,fgk01}.

The improved understanding of the situation for a static spherically
symmetric Black Hole also seems a good basis for extending our result
to include e.g. {}``grey factors'' (ref. 50 in \cite{gkv02}) by admitting
a non-vanishing classical scalar background field, back reaction upon
the metric, and also the consideration of higher orders in the
covariant perturbation series.
\\
\\
\textbf{\large Acknowledgement}
\\
\\
The authors thank especially D. V. Vassilevich (Leipzig) for clarifying
discussions and D. Grumiller for a careful reading of the manuscript.
This work was supported in part by the Austrian Science Foundation (FWF)
Project 14.650-TPH which is gratefully acknowledged.

\newpage
\begin{appendix}

\section{Spherical Reduction}

The spherical reduction procedure is the basis of the
two-dimensional dilaton model considered in
this work. We reproduce it shortly, because some intermediate formulas
are important for the main text. Here we consider a more general
reduction from a $d$-dimensional spacetime $M$ with
spherically-symmetric metric
\begin{equation}
ds^{2}=g_{\alpha\beta}dx^{\alpha}dx^{\beta}-\Phi^{2}(x^{\alpha})
g_{\kappa\lambda}dx^{\kappa}dx^{\lambda}
\end{equation}
to a two-dimensional Lorentz submanifold $L$, spanned by the
coordinates $x^{\alpha}$ (e.g. $t,r$), by reducing out a
$(d-2)$-sphere $S^{d-2}$ with coordinates $x^{\kappa}=\theta,\vp\dots$.
The dilaton field is defined as $X=\Phi^{2}$, $\Phi$ is more
convenient for calculations.
$g_{\alpha\beta}$ is the induced metric on the $L$ and
$g_{\kappa\lambda}$ the one on $S^{d-2}$.
We work in a vielbein basis
in which the line-element can be written as
\begin{equation}
ds^{2}=\eta_{ab}e^{a}\otimes e^{b}-\delta_{ij}e^{i}\otimes e^{j}.
\end{equation}
The $e^r$ form a vielbein basis on $M$. One can define
a vielbein basis on $L$ and $S^{d-2}$ which we denote
by $\tilde{e}^a$ and $\tilde{e}^i$, respectively\footnote{We mark
geometric objects belonging to $L$ or $S^{d-2}$ by a tilde;
tensorial objects on the submanifolds are distinguished
easily by the different indices used.}.
They are related to the $e^r$ by
\begin{eqnarray}
e^{a} &=&\tilde{e}^{a}\,\,,\,\,e^{i}=\Phi\tilde{e}^{i}
\label{vielbein transformation} \\
E_{a} &=&\tilde{E}_{a}\,\,,\,\,E_{i}=\Phi^{-1}\tilde{E}_{i}.
\end{eqnarray}
Further, a (Levi-Civit\'a) spin-connection $\omega^r{}_s$ on $M$
induces connections on the submanifolds:
\begin{equation}
\tilde{\omega}^a{}_b=\omega^a{}_b\,\,,\,\,\tilde{\omega}^i{}_j=\omega^i{}_j.
\end{equation}
The connection on $M$, determined by vanishing
torsion on $M$ and $L$, is given by
\begin{equation}
\omega^r{}_s=\left(\begin{array}{cc}\tilde{\omega}^a{}_b &
(\tilde{E}^a\Phi)\tilde{e}_i\\(\tilde{E}_a\Phi)\tilde{e}^i&
\tilde{\omega}^i{}_j\end{array}\right).\label{con.sph.red.}
\end{equation}
With (\ref{con.sph.red.}) the Riemann tensor on $M$
can be expressed by geometrical objects on $L$ and $S^{d-2}$.
For the scalar curvature one has
\begin{multline}
R^{M}=R^{L}-\frac{1}{\Phi^{2}}\left[R^{S}
+(d-2)(d-3)\left(\tilde{E}_{b}\Phi\right)
\left(\tilde{E}^{b}\Phi\right)\right]\\
-\frac{2}{\Phi}(d-2)\left[ \left( \tilde{E}_{b}\tilde{E}
^{b}\Phi \right) +\left( \tilde{E}^{b}\Phi \right)\tilde{\omega}^{a}{}_b
\left(\tilde{E}_{a}\right) \right]\\
=R^{L}-\frac{\left( d-2\right) \left( d-3\right) }{\Phi ^{2}}
\left[ 1+\tilde{\nabla}_{b}\Phi \tilde{\nabla}^{b}\Phi \right] -2\left( 
\frac{d-2}{\Phi }\right)\tilde{\square}_L\Phi.\label{R-sph.red.}
\end{multline}
In the last line we have inserted the (constant) scalar curvature
of the $(d-2)$-sphere $R^S=\left(d-2\right)\left(d-3\right)$
(the Riemann tensor on the $(d-2)$-sphere is given by $\tilde{R}^i{}_j=
\tilde{e}^{i}\wedge \tilde{e}^{j}$). In this form all quantities
on the r.h.s. live on $L$ as it should be.
By reduction from $d=4$ the Ricci tensor and the scalar curvature become
\begin{eqnarray}
R^M_{ab}&=&R_a(E_b)=R^L_{ab}-2\frac{\tilde{\nabla}_{a}
\tilde{\nabla}_{b}\Phi}{\Phi}=R^L_{ab}+\frac{\tilde{\nabla}_{a}X
\tilde{\nabla}_{b}X}{2X^2}-\frac{\tilde{\nabla}_{a}\tilde{\nabla}_{b}X}{X}\\
R^M&=&R^L-2\frac{1+(\tilde{\n}\Phi)^2}{\Phi^2}
-4\frac{\tilde{\square}\Phi}{\Phi}
=R^L-\frac{2}{X}+\frac{(\tilde{\n}X)^2}{2X^2}-2\frac{\tilde{\square}X}{X}.
\label{R-sph.red.4d}
\end{eqnarray}
In the last equality we have returned to the dilaton field $X=\Phi^2$.
Note that the $\frac{\tilde{\square}X}{X}$-term in the scalar
curvature becomes a surface term in the action when it is multiplied by the
SR measure $\sqrt{-g_M}=X\sqrt{-g_L}$.
For a d'Alembertian on the $d$-dimensional manifold $M$ acting
on a scalar field $S(x^{\alpha})$ which depends only on the
coordinates of $L$ one gets
\begin{multline}
\square S=\eta^{rs}\n_rE_sS=\eta^{ab}\n_aE_bS+\eta^{ij}\n_iE_jS\\
=\tilde{\square}S-\eta^{ij}\omega^a{}_j(E_i)E_aS
=\tilde{\square}S-\eta^{ij}\frac{\tilde{E}^a\Phi}{\Phi}\tilde{E}_aS\\
=\tilde{\square}S+(d-2)\frac{\tilde{\n}^a\Phi\tilde{\n}_aS}{\Phi}
=\tilde{\square}S+\frac{d-2}{2}\frac{\tilde{\n}^aX\tilde{\n}_aS}{X}.
\label{Laplace-sph.red.}
\end{multline}
If $M$ is the four-dimensional Schwarzschild spacetime and
the gauge of the dilaton is fixed as $X=r^2$ this simplifies to
\begin{equation}
\square S=\tilde{\square}S-\frac{2}{r}\Sss\p_rS.
\end{equation}

\section{Green Function Perturbation - Higher Orders}

In this Appendix the second and third order of the
Green function perturbation series (\ref{pert.Green})
are adapted to the case of a 2d
Schwarzschild spacetime with perturbing d'Alembertian (\ref{pert.Laplace}),
taking into consideration the particular boundary conditions
on the half-plane and the time-independence of the involved integrals.
The second order of the perturbation series then reads ($\p''=\p_{r''}$)
\begin{equation}
-\int''[\p''G_0(x,x'')]g(r'')
\p''G_0(x'',x')d^2x'',
\end{equation}
where one $r''$-derivative has been partially integrated and
\begin{equation}
g(r):=\frac{2M}{r}.
\end{equation}
On a two-dimensional Schwarzschild spacetime $\p_r^2g(r)=-R(r)$.
A useful identity can be derived, introducing flat light-cone
derivatives $\p_{\pm}=\p_t\pm\p_r$\,, $\square_0=\p_+\p_-=\p_-\p_+$.
Because of the time-independence one can effectively
set $\square_0=-\p_r^2$ and $\p_+=-\p_-=\p_r$, ($g''=g(r''),
G_0(x,x')=G_{xx'},\square_0G_{xx'}=-\delta^2(x-x')$):
\begin{multline}
0=\int''\square_0''(G_{xx''}G_{x''x'}g'')\\
=-(g+g')G_{xx'}-\int''\left\{(G_{xx''}G_{x''x'})R''
+2g''(\p''G_{xx''})\p''G_{x''x'}\right\}\label{identity}
\end{multline}
By (\ref{identity}) the second order of the perturbation
series (\ref{pert.Green}) can be written in the compact form
\begin{equation}
\frac{1}{2}\left\{[g(r)+g(r')]G_0(x,x')+\int''G_0(x,x'')
G_0(x'',x')R(r'')d^2x''\right\}.\label{pert.sec.order}
\end{equation}
In a similar manner the third order is computed,
\begin{multline}
\int''\int'''G_{xx''}\p''(g''\p''G_{x''x'''})\p'''(g'''\p'''G_{x'''x'})\\
=\int''(\p''G_{xx''})g''\p''\int'''(\p'''G_{x''x'''})g'''\p'''
G_{x'''x'}\\
=\frac{1}{4}\biggm\{[g^2+(g')^2]G_{xx'}+\int''G_{xx''}[\p''(g'')^2]
\p''G_{x''x'}
+[g'g+(g')^2]G_{xx'}\\+g'\int''G_{xx''}G_{x''x'}R''
+\int'''G_{x'''x'}R'''\left([g+g''']G_{xx'''}+\int''G_{xx''}G_{x''x'''}R''
\right)\biggm\},
\end{multline}
where again the identity (\ref{identity}) has been used.
Finally, we write down the third order of the perturbation
series in a compact form, replacing $\p_rg^2(r)=2R(r)$:
\begin{multline}
\frac{1}{4}\biggm\{[g^2(r)+g(r)g(r')+2g^2(r')]G_0(x,x')\\
+\int''G_0(x,x'')R(r'')[2\cdot\p_{r''}+g(r')]G_0(x'',x')d^2x''\\
+\int'''G_0(x''',x')R(r''')\\
\cdot\left([g(r)+g(r''')]G_0(x,x''')+\int''G_0(x,x'')G_0(x'',x''')
R(r'')d^2x''\right)d^2x'''\biggm\}.\label{pert.thi.order}
\end{multline}

\section{Non-Conservation Equation}

Diffeomorphism invariance
of the generating functional implies a non-conser\-vation
equation for the renormalized expectation value of the EM tensor
when a dilaton field is present, resulting from the non-minimal
coupling of the scalar field in $d=2$ (cf. (\ref{2daction})).
A $2d$ diffeomorphism transformation $\delta_{\xi}g^{\alpha\beta}
=-(\n^{\alpha}\xi^{\beta}+\n^{\beta}\xi^{\alpha}),\delta_{\xi}X
=\xi^{\alpha}\p_{\alpha}X,\delta_{\xi}S=\xi^{\alpha}\p_{\alpha}S$
applied to the effective action $W[g]=-i\ln Z[g]$ yields the
\emph{quantum} non-conservation equation which has not been
checked in the previous literature:
\small
\begin{eqnarray}
0&=&\delta_{\xi}W[g]=\frac{-i\delta_{\xi}Z[g]}{Z[g]}=\frac{-i{\cal N}}
{Z[g]}\delta_{\xi}\int{\cal D}(\sqrt[4]{-g}S)
\,\,e^{i\int_L\frac{X}{2}\left[(\n S)^2
-m^2S^2\right]\sqrt{-g}d^2x}\nonumber\\
&=&\frac{-i{\cal N}}{Z[g]}\int{\cal D}(\sqrt[4]{-g}S)\int_L^y\left[
\delta_{\xi}g^{\alpha\beta}\frac{\delta}{\delta g^{\alpha\beta}}+
\delta_{\xi}X\frac{\delta}{\delta X}+\delta_{\xi}S\frac{\delta}{\delta S}
\right]e^{L_{dil}^m}\,d^2y\nonumber\\
&=&\frac{-i{\cal N}}{Z[g]}\int{\cal D}(\sqrt[4]{-g}S)\,\,
e^{L_{dil}^m}\int_L^y\bigg\{
-(\n^{\alpha}\xi^{\beta}+\n^{\beta}\xi^{\alpha})i\frac{\sqrt{-g}}{2}
T_{\alpha\beta}\nonumber\\
&&+\xi^{\alpha}\p_{\alpha}X\frac{i\sqrt{-g}}{2}
\left[(\n S)^2-m^2S^2\right]\bigg\}\,d^2y\nonumber\\
&&+\frac{i{\cal N}}{Z[g]}\int{\cal D}(\sqrt[4]{-g}S)\,\,
e^{L_{dil}^m}\int_L^y\frac{\delta}{\delta S}\delta_{\xi}S\,d^2y\nonumber\\
&=&\frac{{\cal N}}{Z[g]}\int{\cal D}(\sqrt[4]{-g}S)\,\,
e^{L_{dil}^m}\int_L^y\xi^{\alpha}
\bigg\{\n^{\beta}T_{\alpha\beta}+\frac{\p_{\alpha}X}{2}
\left[(\n S)^2-m^2S^2\right]\nonumber\\
&&+\frac{i}{\sqrt{-g}}\lim_{x\to y}\p_{\alpha}\delta(x-y)
\bigg\}\sqrt{-g}d^2y\nonumber\\
&=&\int_L^y\xi^{\alpha}
\bigg\{\n^{\beta}\E{T_{\alpha\beta}}+\frac{\p_{\alpha}X}{2}
\left[\E{(\n S)^2-m^2S^2}\right]\nonumber\\
&&+\frac{i}{\sqrt{-g}}\lim_{x\to y}\p_{\alpha}\delta(x-y)
\bigg\}\sqrt{-g}d^2y
\end{eqnarray}
\normalsize
In the step from the second to the fifth line
a partial integration in the path integral has been performed.
The divergence $\delta(0)$ represents the infinite zero-point energy
of the quantized scalar field $S$. Defining the renormalized
EM tensor $\E{T_{\alpha\beta}}_{ren}:=\E{T_{\alpha\beta}}
+\frac{ig_{\alpha\beta}}{\sqrt{-g}}\lim_{x\to y}\delta(x-y)$ we obtain
indeed (\ref{EM.cons.quant.dil}), as proposed.

\section{Heat Kernel Integrals}

The leading divergent ($T\to\infty,\ve\to0$) terms of the expansions
(\ref{div1}-\ref{div5}), appearing in the trace of the heat kernel
(\ref{nonlocal.effective.action}) can be computed analytically.
For instance (\ref{div2})
only has an UV divergence, hence we cutoff the
$\tau$-integration at the lower boundary by $\ve>0$.
As a first step we expand the
regularized integrand in powers of $s$, carry out the differentiation
for $s$ and then set it zero:
\begin{multline}
\frac{d}{ds}\left\{\frac{1}{\Gamma(s)}\int_{\ve}^{\infty}\tau^{s-1}
d\tau\right\}\biggm|_{s=0}\\
=\frac{d}{ds}\left\{[s+s^2\gamma_E+O(s^3)]
\int_{\ve}^{\infty}[\tau^{-2}+s\tau^{-2}\ln\tau+O(s^2)]
d\tau\right\}\biggm|_{s=0}\\
=\int_{\ve}^{\infty}\tau^{-2}d\tau=\frac{1}{\ve}.
\end{multline}
The most problematic term is (\ref{div3}). Introduction of a cutoff
at infinity and differentiation for $s$ leads to (using (\ref{f2}):
\begin{multline}
\frac{d}{ds}\left\{\frac{1}{\Gamma(s)}\int_0^{T}\tau^{s}
f(-\tau\triangle)d\tau\right\}\biggm|_{s=0}=\int_0^1\int_0^T
e^{-\tau a(1-a)(-\triangle)}d\tau da\\
=\int_0^1\frac{e^{-T a(1-a)(-\triangle)}-1}{a(1-a)(-\triangle)}da
=\frac{4}{-\triangle}\int_0^1\frac{e^{(z^2-1)\frac{T(-\triangle)}{4}}-1}
{z^2-1}dz
\end{multline}
This integral cannot be solved analytically, however, it
can be compared to
\begin{multline}
I=\int_0^1\int_0^Te^{(z-1)\frac{\tau(-\triangle)}{4}}d\tau dz
=\frac{4}{-\triangle}\int_0^1\frac{e^{(z-1)\frac{T(-\triangle)}{4}}-1}
{z-1}dz\\
=\frac{4}{-\triangle}\left\{\ln\left[\frac{T(-\triangle)}{4}
\right]+\gamma_E\right\}\label{aux.integral}.
\end{multline}
This is twice the result conjectured already at the r.h.s. of
(\ref{div3}) apart from an additive constant
$\frac{4\ln4}{\triangle}$. Hence, it remains to show that the
difference between $I/2$ and of the original
expression (l.h.s. of (\ref{div3})) converges to (one half of)
that constant for large
regularisation parameter $T$. First we perform a substitution of the
integration variable $z\to z^2$ in (\ref{aux.integral}):
\begin{equation}
\frac{I}{2}=\int_0^T\int_0^1z\cdot e^{(z^2-1)
\frac{\tau(-\triangle)}{4}}dzd\tau
\end{equation}
This difference then becomes
\begin{multline}
\int_0^1\int_0^Te^{(z^2-1)\frac{\tau(-\triangle)}{4}}(z-1)d\tau dz=
\frac{4}{-\triangle}\int_0^1\frac{e^{(z^2-1)\frac{T(-\triangle)}{4}}-1}
{z+1}dz\\
\stackrel{T\to\infty}{\to}\frac{4}{\triangle}\int_0^1\frac{1}{z+1}dz
=\frac{4\ln2}{\triangle}=\frac{2\ln4}{\triangle}
\end{multline}
which proves (\ref{div3}) to be the correct
result for large $T$. We note that the limit\newline
$\lim_{T\to\infty}e^{(z^2-1)\frac{T(-\triangle)}{4}}=0$
could be performed because $z^2<1$ for all $z$ except for $z=1$.
But at that value the integrand vanishes altogether.

The computation of the remaining expressions is tedious but straightforward.

\end{appendix}

\end{document}